\documentstyle[11pt,psfig,aaspp4]{article}

\def\kms{km~s$^{-1}$}
\def\etal{{\it et al.}~}

\def\Nnod{157~}
\def\Ndet{587~}
\def\Ntot{744~}
\def\W50{W$_{50}$}

\font\eightrm=cmr8
\font\eightbf=cmbx8
\font\eightit=cmti8
\font\eightsl=cmsl8
\font\eightmus=cmmi8
\def\smalltype{\let\rm=\eightrm \let\bf=\eightbf
  \let\it=\eightit \let\sl=\eightsl \let\mus=\eightmus
  \baselineskip=9.5pt minus .75pt
  \rm}

\begin{document}

\title{Spectroscopy of Edge--on Spirals}
\author{Riccardo Giovanelli and Eric Avera}
\affil{Center for Radiophysics and Space Research}
\affil{and National Astronomy and Ionosphere Center\footnote{The National
Astronomy and Ionosphere Center is operated by Cornell
University under a cooperative agreement with the National Science Foundation.}}
\affil{ Cornell University, Ithaca, NY 14853}
\affil{riccardo@astrosun.tn.cornell.edu}

\author{Igor D. Karachentsev}
\affil{Special Astrophysical Observatory, Russian Academy of Sciences}
\affil{ikar@smile.sao.ru}

\hsize 6.2 truein

\begin{abstract}
We present HI line observations of \Ntot edge--on spiral galaxies, extracted
from the {\it Flat Galaxy Catalog} of Karachentsev \etal (1993). Fluxes, systemic 
velocities and line widths are given for \Ndet detected galaxies, as well as 
search parameters for \Nnod undetected systems.
Widths are corrected for instrumental broadening,
smoothing, signal--to--noise and profile shape, and
an estimate of the error on the width is given. 
When corrected for turbulent broadening and 
inclination angle of the disks, the velocity widths presented here
can provide the appropriate line width parameter needed to derive
distances via the Tully--Fisher relation.
\end{abstract}

\section{Introduction}

The neutral hydrogen 21 cm line is a useful tool both for
the quantification of atomic gas content and for the measurement of
distances, either directly from the redshift via Hubble's law,
or indirectly by applying the Tully--Fisher (1977; TF) relation. 
The study of deviations from Hubble flow has received substantial
impetus from the accumulation of large redshift--independent distance
data bases, especially for spiral galaxies, and a detailed picture of the
distribution of mass, luminous and dark, in the local Universe is
starting to emerge (Dekel 1994; Strauss and Willick 1995; da Costa
\etal 1996; Giovanelli 1997a). Measurements of distance by means of the
TF method (or of analogous relations for spheroidal galaxies) are however
notoriously noisy, yielding estimates of peculiar velocity with accuracy
of about 15\% of the galaxy recessional velocity. As a result, maps of
the peculiar velocity field at redshifts $cz$ exceeding a few thousand
\kms require large scale volume averaging, and thus sampling densities
which rapidly approach observationally prohibitive levels. It is then
desirable to explore avenues that may lead to obtaining more accurate 
distance estimation techniques. In 1989, Karachentsev suggested the use
of a variant on the standard TF relation. Using a small data base of
edge--on spiral systems with accurate photometric and spectroscopic
measurements, he showed that a tight correlation may exist between
the disk scale length, the central disk surface brightness and the 
rotational velocity, a variant on the standard
TF relation between the absolute magnitude and the rotational velocity.

It is well--known how inclination--dependent corrections affect the error 
budget of the TF relation: while for edge--on systems the corrections necessary
to recover the rotational velocity from the 
observed velocity width are smallest, and thus the associated errors
introduced by such corrections are minimized, internal extinction corrections
are largest, and the associated uncertainty is in turn maximized. The opposite occurs for 
more nearly face--on systems. Since even in the near infrared, such as 
the I band, extinction corrections are still quite important (Giovanelli
\etal 1994, 1995), variants of the TF relation that are less sensitive
to magnitude corrections become attractive. The one proposed by
Karachentsev (1989), and reiterated by Chiba and Yoshii (1996), motivated 
the preparation of a catalog of edge--on spiral systems (Karachentsev \etal 1993).
These objects, by virtue of their selection by thresholding extreme values 
of the axial ratio, are most likely to be nearly bulge--free disk galaxies. 
In that case, it is presumed that the measurement of disk scaling properties 
would be facilitated, and that they would be ideally suited for the applications 
of the mentioned TF variant relation. In a recent review of applications of
the TF relation and its variants, Giovanelli (1997b) has pointed out possible
limitations in the approach proposed by Karachentsev (1989) and Chiba and
Yoshii (1996), which result from uncertainties associated with the 
measurement of central disk surface brightnesses and disk scale lengths
(see also Knapen \& van der Kruit 1992 and Byun \etal 1994). While the
highly desirable discovery of a superior form of the TF technique awaits
confirmation, nearly edge--on spirals remain primary candidates for the 
application of most of its variants.

This work presents spectroscopic observations of a set of nearly edge--on galaxies
extracted from the Karachentsev \etal (1993) catalog, carried out with the Arecibo
telescope. The sample contains a number of objects observed previously for
other purposes, which have been added to the present compilation.
Details of the observations are briefly summarized and the parameters of
the processed data are presented in a form that makes them suitable for
large scale structure and distance determination studies, as well as for
studies concerned with the global properties of this most interesting
category of galaxies.

\section{Observations}

New HI line observations of cluster galaxies were conducted using the 
Arecibo 305m telescope during a number of separate observing periods between
mid--1989 and early 1994. In this section, a summary of the details
of those observations is presented. A number of objects were observed
previously, and those observations have been reprocessed for consistency
in the final output of spectroscopic parameters.

The observing target list was obtained by selecting galaxies
from the FGC and FGCA catalogs of Karachentsev \etal (1993), which lare located
between $0^\circ$ and $38^\circ$ in declination, and thus reachable by the
Arecibo telescope. 
All but six of the galaxies were observed at Arecibo; the remainder were
observed either with the Nan\c cay telescope of the Observatory of Paris
or with the 92--meter and 42--meter telescopes of the National Radio Astronomy 
Observatory.  

The observational setup used for the new Arecibo observations is essentially
that described in Giovanelli \& Haynes (1989) and by Haynes \etal (1997), and
we do not further expand on them here. Haynes \etal (1997) also describe in
detail the growing impact of radio frequency interference (rfi) on extragalactic
HI observations. Rfi affects with equal severity this set of data as the one
presented in that reference. 

In addition to the new observations, most HI line spectra obtained
by us and our collaborators in the last decade are available in a 
digital archive. In order to achieve homogeneous velocity width 
measurements, all spectra presented here have been reanalyzed using the Arecibo
ANALYZ-Galpac package following a uniform processing path.
Final parameters derived from the reanalysis include fluxes,
systemic velocities, and velocity widths for detected galaxies
and limits on the rms noise per channel for undetected ones.

Of particular importance to the use of these spectra for
TF applications, velocity widths for all galaxies have been 
measured using a new algorithm specifically designed to give both
a more robust width measurement and a reliable quantitative assesment
of the width error. The adopted algorithm measures 
\W50, defined as the full width across 
the profile measured at a level of 50$\%$ of each horn, where the
appropriate level is identified by fitting a polynomial between the levels of
20$\%$ and 80$\%$ of the respective horn on each side of the profile.
The velocity corresponding to the 50$\%$ level is estimated from the fitted 
polynomial rather than from the observed spectrum. In addition, we present a
standard determination of the width measured at a level equal to
20\% of the flux at the profile horns, for comparison with other
work that uses that width estimator. This measurement is however far
more susceptible to spectrum quality (i.e. signal--to--noise ratio), and 
the associated errors are typically larger than those on \W50.

A correction to the observed profile width for instrumental effects 
must account for the broadening of the profile produced by noise, 
smoothing and the non--zero spectrometer channel width. While smoothing 
the spectrum usually allows a better determination of the central velocity,
it also increases systematically the measured width. In the current data set 
we have applied the simplest Hanning smoothing (convolution with a
[0.25,0.5,0.25] three--point function) in processing the available data, 
whether newly obtained or previously published, whenever the signal--to--noise
ratio of the data was sufficiently high. This form of smoothing is generally
necessary in order to remove the $(\sin x)/x$ ringing introduced by narrow rfi
features, therefore even very high signal--to--noise spectra were thus smoothed. 
In cases of very low signal--to--noise spectra, however, it was necessary to 
apply more robust smoothing, in order to reduce the noise to a tolerable level;
in those cases, two successive smoothing steps were taken: first the spectrum
was convolved with a 3--channel wide boxcar function, followed by Hanning as
described above. To the measured widths, a statistically derived correction was 
later applied, which takes into consideration the broadening produced by
the smoothing process. A record of the parameters required for the estimate of 
corrections to the observed \W50 is maintained through our analysis process. 
The statistical derivation of the correction recipes will be presented elsewhere 
(Haynes \etal, in preparation). 

Many of the observations reported here have been made for other purposes,
besides those indicated in this paper. The largest overlap, of 153 objects, corresponds with an all--sky
program of I--band TF distances of field Sbc and Sc galaxies (Giovanelli \etal 1997b).
Entries for those galaxies are reproduced in this paper for completeness,
and flagged accordingly.

\subsection{Results for Detections}

We present in Table 1 the results of the processing of the spectra
for \Ndet galaxies. Figure 1 shows a sample of the HI spectra, including 
those
of the first 16 galaxies listed in the table. The smoothed curve
in each profile is the polynomial baseline subtracted from each
spectrum before derivation of the emission properties. Strong,
typically narrow band rfi is present in many of the spectra.
A digital version of Table 1 and a copy of the complete Fig. 1 
can be obtained upon request from RG.

Details of the entries in Table 1 are as follows:

\noindent 
Col 1: Entry number in the UGC (Nilson 1973), where applicable, or else
in our private database, referred to as the Arecibo General Catalog (AGC).

\noindent 
Col 2: FGC or FGC Addenda designation.

\noindent
Cols. 3 and 4: Right Ascension and Declination in the 1950.0 epoch,
either from the literature or measured by us on the POSS--I.
Typically, the listed positions have 15" accuracy. Although more
accurate positions are now easily available, they were not at the
time of the observations, and the positions actually used for the
observations are reported here, unless otherwise indicated in the notes.

\noindent
Col. 5: The blue major and minor diameters, a $\times$ b, in 
arcminutes, from the FGC as estimated on blue POSS--1 prints.

\noindent
Col. 6: The observed integrated 21 cm HI line flux $S = \int S_\nu d\nu$
in Jy-\kms.

\noindent
Col. 7: The corrected integrated 21 cm HI line flux $S_c$, also in
Jy-\kms, after corrections applied for pointing offsets (Arecibo only)
and source extent following Haynes \& Giovanelli (1984). For galaxies
in this list for which UGC blue sizes are available, we use those for the
estimate of the flux correction, as the relationship between HI size and
optical size was obtained using that type of data; for other galaxies, we use
the FGC size to estimate the flux correction.

\noindent
Col. 8: The rms noise per channel of the spectrum, rms, in mJy.

\noindent
Col. 9: The emission profile signal to noise ratio, snr, taken as the
ratio of the peak flux to the rms noise, after smoothing as specified
in col. 15 .

\noindent
Col. 10: The heliocentric velocity V$_\odot$, in \kms, of the HI line
signal, which is the midpoint of the profile at the 50$\%$ level also
used to measure the width (see below). The error on the velocity is 
typically about 1.4 times smaller than that on the velocity width reported
in col. 14.

\noindent
Col. 11: The full velocity width W$_{50}$ measured at the 50\% level, in \kms, 
of the HI line, uncorrected for redshift of other effects. The measurement 
algorithm is discussed in Giovanelli \etal~ (1997a).

\noindent
Col. 12: The full velocity width W$_{20}$ measured at the 20\% level, in \kms, 
of the HI line, uncorrected for redshift of other effects. This width is of
lower quality than \W50.

\noindent
Col. 13: Corrected 50\% velocity width W$_{50,c}$, in \kms. The correction 
accounts for redshift stretch, instrumental broadening and smoothing during the 
processing phase. Note that this width is not rectified for either the viewing 
angle of the disk to the line of sight or for turbulent broadening.

\noindent
Col. 14: The estimated error on W$_{50,c}$, $\epsilon_w$, in \kms, taken to be
the sum in quadrature of the measurement error and the estimated error in the
instrumental and processing broadening corrections. The estimated error on 
W$_{20}$ is typically larger than that for W$_{50,c}$, as it tends to be more 
affected by spectrum noise. A careful appraisal of that error was not carried
out and thus such number is not reported. 

 \noindent
Col. 15: A series of codes indicating the mode of the spectrometer configuration, 
smoothing applied, width quality and data source.

The first code refers to the telescope/spectrometer configuration of the final 
spectrum. Codes {\it a} and {\it o} refer to the Arecibo telescope, as follows:
{\it a} : 20 MHz total bandwidth over 512 spectral channels; {\it o} : 10 MHz/252 
channels; code {\it g} : 10 MHz/192 channels, data taken with the 
92--meter telescope in Green Bank; code {\it b} : 10 MHz/512 channels,  
data taken with the 43--meter telescope in Green Bank and, finally,  
code {\it n} : 6.4 MHz/256 channels , data taken with the Nan\c cay telescope.

The second letter is the smoothing code: H refers to Hanning only; 
B refers to convolution with 3 channel boxcar followed by Hanning.

The third code is a qualitative assessment of the quality of the profile
for TF applications: G = good; F = fair; M = marginal detection; C = confused;
S = single peak and P = completely unfit for TF use.
A designation of marginal detection is given to those cases in which the signal
is of very poor signal--to--noise and has not been verified through adequate 
reobservation. 

A final $*$ in this column indicates the presence of comments
in the following notes to Table 1, while a final $s$ signifies that
the galaxy is also part of the {\it Sc} galaxy project of Giovanelli
\etal (1997b).

\subsubsection{Notes to Table 1}
{\smalltype
\noindent 100740=F36: blended or perturbed spectrum; possibly more than one gal in beam;
spiral comp 2.5' S and UGC214 at 7' and vhel=5379.

\noindent 819=F143: velocity of Schneider \etal (1990, ApJS 72, 245) not confirmed and prob. in error;
our observations refer to two different epochs, in agreement with each other.

\noindent 1773=F275: prob. interacting system; blend?

\noindent 2097=FA40: marginal detection of second, weaker feature at vhel=3092,W=271
needs confirmation.

\noindent 2159=F328: v. asymm profile.

\noindent 2201=F337: U2204=N1067 and U2203=N1066 also in the beam; U2204 also detected; no
confusion.

\noindent 2344=F350: very marginal detection; parms highly uncertain; reading tea--leaves
would suggest vhel$\sim$14730 as next likelier alternative.

\noindent 130395=F435: other feature also detected and confirmed at vhel=8780; close to
GPS rfi, but unlikely to be caused by it; some very low SB features within beam,
visible on blue PSS.

\noindent 130382=F464: 130383 also in beam, separated by 0.3', both Sc edge--on; two
features in HI spectrum, respectively at vhel=6674, W=215 and vhel=5801, W=259;
no blend. Identification of optical image with feature in radio spectrum 
impossible: tentative assignment arbitrary.

\noindent 130383: see 130382

\noindent 3005=F473: U3004 also in the beam and in the HI spectrum at vhel=3573; no blend.

\noindent 180111=F726: two galaxies in beam; prob. blend. Caution with parms.

\noindent 180620: serendipitously detected in ``off'' of 180611.

\noindent 180605=F765: comp. in beam at 1.0' S; prob. blend.

\noindent 180606=F772: comp. in beam at 0.8' W; prob. non--interfering detection at
vhel=6884, W50=321, flux integral 0.39.

\noindent 4591=F793: in tight group with U4589, U4595 and sev. other fainter 
spirals in vicinity. HI spectrum is blend: parms. of poor reliability for TF.

\noindent 190712=FA110: v. asymmetric profile, possible contamination by rfi;
unreliable for TF use.

\noindent 190744=F913: strong standing waves in spectrum; S/N poor.Caution.

\noindent 190716=F980: two features in HI spectrum, respectively at vhel=8016,
W=386, and vhel=5700,W=207; smaller galaxy (190717) also in beam. 
Match of image and spectrum somewhat arbitrary.

\noindent 6306=FA150: U6305 in beam; no confusion.

\noindent 6428=FA152: 21cm observations made with the 300--foot telescope.

\noindent 210559=F1287: asymm. profile, possibly confused.

\noindent 6715=F1300: prob. interacting with comp. at 2'; blended profile.

\noindent 211029=F1319: prob. confused with emission from sev. other 
objects in beam.

\noindent 7510=F1429: blended with galactic HI, gas deficient; width unreliable.

\noindent 7772=F1471: resolved,so that single beam doesn't collect full
flux; high parm. unreliability, not reflected in measurement 
error.

\noindent 9000=F1711: very asymm. profile: blend with U9001. Parms. unreliable.

\noindent 9345=F1762: blend with U9346; parms. unreliable for TF work.

\noindent 241183=F1832: feature detected at vhel=9313, W=63, with no obvious
optical counterpart.

\noindent 250015=F1846: sev. comps. in vicinity: blend?

\noindent 9760=F1863: feature detected at vhel=1858, W=35, with no obvious
optical counterpart.

\noindent 9977=F1935: 21cm observations made with the Nan\c cay telescope.

\noindent 10027=F1949: asymm. spectrum.

\noindent 251587=F1965: rugged, asymm. profile; width uncertain.

\noindent 251333=F1968: blended profile; Sc comp., 0.6'x0.45', at 1.6' also in 
beam.

\noindent 10219=F1989: also observed at Nan\c cay, with flux=13.88, vhel=1368,
W=209.

\noindent 10232=F1995: observed with telescope pointing at RA=160732.3; because
of large position error, flux is underestimated.

\noindent 10274=F2001: 21cm observations made with the 300--foot telescope.

\noindent 10625=F2091: 21cm observations made with the 300--foot telescope. 
Gaussian profile.

\noindent 10716=F2111: feature detected at vhel=9426, W=60 partly blended with
galaxy emission.

\noindent 11093=F2176: 21cm observations made with the 140--foot telescope. 

\noindent 11132=FA253: 21cm observations made with the 300--foot telescope.

\noindent 280138=F2187: marginal det. on edge of spectrum; parms uncertain; needs 
confirmation.

\noindent 11142=F2188: 280139=F2189 at 5'; no evidence of confusion in 
spectrum.

\noindent 280139=F2189: detected on very edge of bandpass; parms. need confirmation. 

\noindent 12253=F2433: observed with beam centered 2.5' off in declination; parms
uncertain, esp. flux.

\noindent 12714=F2520: F2519 at vhel=5081 also in the beam at 2'; no confusion, as
flux of F2519 is barely detected when pointin on it.

\noindent 331175=F2639: detected serendipitously galaxy at 2352388+084709, in ``off'' spectrum of this observation: vhel=5263$\pm$3, W50=163.
}

\subsection{Results for Non--detections}

In addition to the detected objects, \Nnod galaxies were observed but
not detected in the 21 cm line. In some cases, optical velocities are
available and hence the rms noise per channel can be used to give an 
estimate of the upper limit to the HI mass. In other cases, where the redshift
is not known, it is quite likely that the galaxy lies outside the velocity range
covered by the search mode strategy. 

A summary of the observations of undetected galaxies is presented in Table 2.
Columns 1 through 6 list the FGC number, 1950.0 coordinates, blue diameters
and UGC identification. In column 7, a graphic representation of the heliocentric 
velocity search range is given. The typical r.m.s. noise in the
spectra is approximately 1.0 mJy, after 3--channel boxcar and hanning smoothing. All the
reported non detections were obtained in the mode identified by the letter ``a''
in the description of col. 15 of table 1.

\acknowledgements

It is always a pleasure to acknowledge the skill, dedication and 
assistance of the Arecibo telescope operators. One of the authors
(EA) took part in some of the observing runs and contributed significant
input in the preparation of this work. It was however impractical
for him to see a completed draft of the paper and he should therefore
not be considered responsible for any possible errors in the
presentation, for which the senior authors take full responsibility.
This research was supported by NSF grants AST94--20505 to RG. This 
study has made use of the NASA/IPAC Extragalactic Database (NED)
which is operated by the Jet Propulsion Laboratory, California Institute   
of Technology, under contract with the National Aeronautics and Space      
Administration.                                                            
 

\center{Figure Captions}

\figcaption[fig1] {A sample montage of spectral profiles for the
first 16 galaxies included in Table 1. The x-axis is heliocentric
velocity in \kms; the y-axis is flux in mJy. Instrumental baselines
have not been removed from the data, but are indicated as smoothed
curves superimposed to the data. \label{fig:fig1}}

\end{document}


